\documentclass[aps,prl,twocolumn,floatfix,superscriptaddress,notitlepage]{revtex4-1}
\usepackage{tikz}

\usepackage{pgfplots}
\usetikzlibrary{shadows.blur,arrows.meta}

\usepackage{amsmath}
\usepackage{amssymb}
\usepackage{gensymb}
\usepackage{bm}
\usepackage{graphicx}
\usepackage{float}
\usepackage{placeins}
\usepackage{braket}
\usepackage{bbold}
\usepackage[colorlinks, linkcolor=blue, citecolor=blue, urlcolor=blue, breaklinks=red]{hyperref}

\newcommand{\mi}{{\rm i}}

\usepackage{dsfont}

\begin{document}

\title{Photon blockade with ground-state neutral atoms}
\author{A. Cidrim}
\affiliation{Departamento de F\'isica, Universidade Federal de S\~{a}o Carlos, 
Rod.~Washington Lu\'is, km 235 - SP-310, 13565-905 S\~{a}o Carlos, SP, Brazil}
\author{T. S. do  Espirito  Santo}
\affiliation{Instituto de F\'isica de S\~{a}o Carlos, Universidade de S\~{a}o Paulo - 13560-970 S\~{a}o Carlos, SP, Brazil}
\author{J. Schachenmayer}
\affiliation{IPCMS (UMR 7504) and ISIS (UMR 7006), Universit\'e de Strasbourg, CNRS, 67000 Strasbourg, France}
\author{R. Kaiser}
\affiliation{Universit\'e de C\^ote d'Azur, CNRS, INPHYNI, France}
\author{R. Bachelard}
\affiliation{Departamento de F\'isica, Universidade Federal de S\~{a}o Carlos, 
Rod.~Washington Lu\'is, km 235 - SP-310, 13565-905 S\~{a}o Carlos, SP, Brazil}

\begin{abstract}
We show that induced dipole-dipole interactions allow for photon blockade in subwavelength ensembles of two-level, ground-state neutral atoms. Our protocol relies on the energy shift of the single-excitation, superradiant state of $N$ atoms, which can be engineered to yield an effective two-level system. A coherent pump induces Rabi oscillation between the ground state and a collective bright state, with at most a single excitation shared among all atoms. The possibility of using clock transitions that are long-lived and relatively robust against stray fields, alongside new prospects on experiments with subwavelength lattices, makes our proposal a promising alternative for quantum information protocols. 
\end{abstract}

\date{\today}

\maketitle

\begin{figure}[t!]
\centering
\includegraphics[width=\columnwidth]{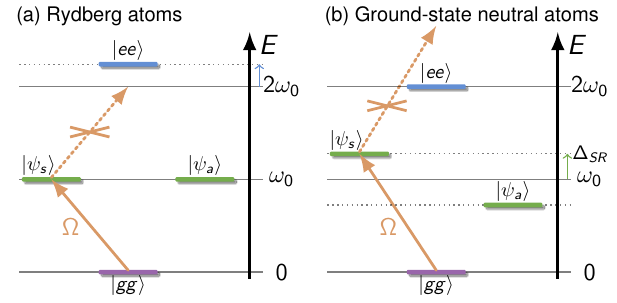} 
\caption{Transition scheme for a pair of atoms (a) in a Rydberg configuration, where the energy shift occurs for the doubly-excited state, the symmetric state is directly addressed using the atomic transition wavelength; (b) in a configuration with strong induced dipole-dipole interactions, the energy of symmetric and antisymmetric single-excitation states shifts oppositely: in this case, to target the symmetric state, the pump is detuned by the interaction-induced shift $\Delta_{SR}$.
\label{Fig:scheme}}
\end{figure}

Photon-induced blockade is a mechanism where multi-excitation states become marginally populated owing to a non-linearity in the excitation spectrum, e.g.~due to energy shifts caused by interactions. In atomic systems, it has been achieved up to date using Rydberg atoms, excited to highly energetic levels, with a principal quantum number of several tens \cite{Saffman2010,Saffman2016,Omran2019,Zhang2020}. Crucially, such setups rely on huge attainable interaction strengths among Rydberg excited atoms, which can be many orders of magnitude larger than other typical interaction strengths (e.g.~involving van der Waals or magnetic dipole-dipole forces) of the ground states. The resulting interaction translates into a strong repulsion between two Rydberg atoms, which shifts the multi-excitation states in energy (see Fig.~\ref{Fig:scheme}a) and can block the presence of more than one excitation inside the so-called blockade radius. Through this mechanism, it is thus possible to address a single-excitation ``super-atom'' state of $N$ atoms (typically a symmetric state). This allows one to simulate an effective two-level system, consisting of a collective excited state of the form $\ket{\psi_s}=(1/\sqrt{N})\sum_{i=1}^{N} e^{i\mathbf{k} \cdot \mathbf{r}_i}\ket{g...ge_ig...g}$ (with $\mathbf{k}$ the pump wavevector and $\mathbf{r}_i$ the position of atom $i$) and a ground state $\ket{\psi_g}=(1/\sqrt{N})\ket{g...g}$. The coupling to the pump with single-atom Rabi frequency $\Omega$ results in the collective Rabi oscillations at frequency $\sqrt{N}\Omega$. They are the hallmark of photon blockade, as they provide a direct observable of the collective state, and serve as a witness of the $N$-atom entanglement. Indeed, starting from ground-state atoms and switching off the pump after a quarter of the collective oscillation fully populates the entangled state $\ket{\psi_s}$. Other signatures, possibly less easily accessible experimentally, are the excitation number of the system, as well as the correlation functions between neighbouring atoms, that shows the suppression of two excitations to be within the same blockade radius. This mechanism makes experiments with Rydberg states of neutral atoms attractive platforms to implement quantum information protocols \cite{Saffman2010,Saffman2016}.

While Rydberg physics can be ideal for the generation of appreciable quantum correlations and entanglement, there can be several downsides of using such systems, which usually stem from challenging experimental control levels. For instance, these atomic states are extremely sensitive to stray electric and magnetic fields due to their large dipole moments \cite{Saffman2005,Arias2019}. Additionally, direct, single-photon excitation from an electronic ground state to a Rydberg state is typically very difficult experimentally due to requirements for short-wavelength lasers~\cite{Takei2016}. Instead, their excitation generally involves a two-photon process, which in turn induces losses during the commonly used stimulated Raman adiabatic passage. Furthermore, it has been shown that the dense spectrum of nearby Rydberg states might severely shorten the time available for coherent manipulation of the Rydberg atoms, implying restrictions to protocols that involve dressing states, for instance \cite{Goldschmidt2016}.

Here we consider instead the case of ground-state neutral atoms that interact solely through induced dipole-dipole interactions and have a simple two-level internal structure. While Rydberg states result in effectively micrometer-sized atoms, in order to reach strong enough interaction strengths in our system, we consider subwavelength samples. Such systems have recently drawn considerable attention, as dipole-dipole interactions at short distances can generate spin squeezing \cite{Qu2019} and also extraordinary subradiance, allowing for long-lived entangled states that may eventually behave as quantum memories \cite{Guerin2016,Asenjo2017, Asenjo2019, Moreno2019, Needham2019, Ballantine2019,Guimond2019}. Furthermore, recent proposals based on stroboscopic techniques are exploring new ways to overcome the diffraction limit and turning subwavelength arrays into reality \cite{Subhankar2019,Tsui2019}. The advantage of choosing schemes with ground-state neutral atoms over Rydberg-based protocols is the fact that one could benefit from long-lived transitions currently used in highly-controllable atomic clock experiments, which are also less sensitive to stray fields, increasing the robustness for coherent control. 

Although our proposal presents similar physics to the Rydberg blockade, it differs in the fact that the shift in energy occurs not on a highly excited state, but rather on the single-excitation states. The simple case for $N=2$ is depicted in Fig.~\ref{Fig:scheme}b and contrasted with the Rydberg counterpart. Here we address the detuned symmetric state by setting the pump frequency to $\omega_0+\Delta_{SR}$, with $\omega_0$ the bare single-atom transition frequency. This frequency shift $\Delta_{SR}$ prevents populating the two-excitation state, inducing a blockade mechanism.

\begin{figure*}[t!]
\centering
\includegraphics[width=\textwidth]{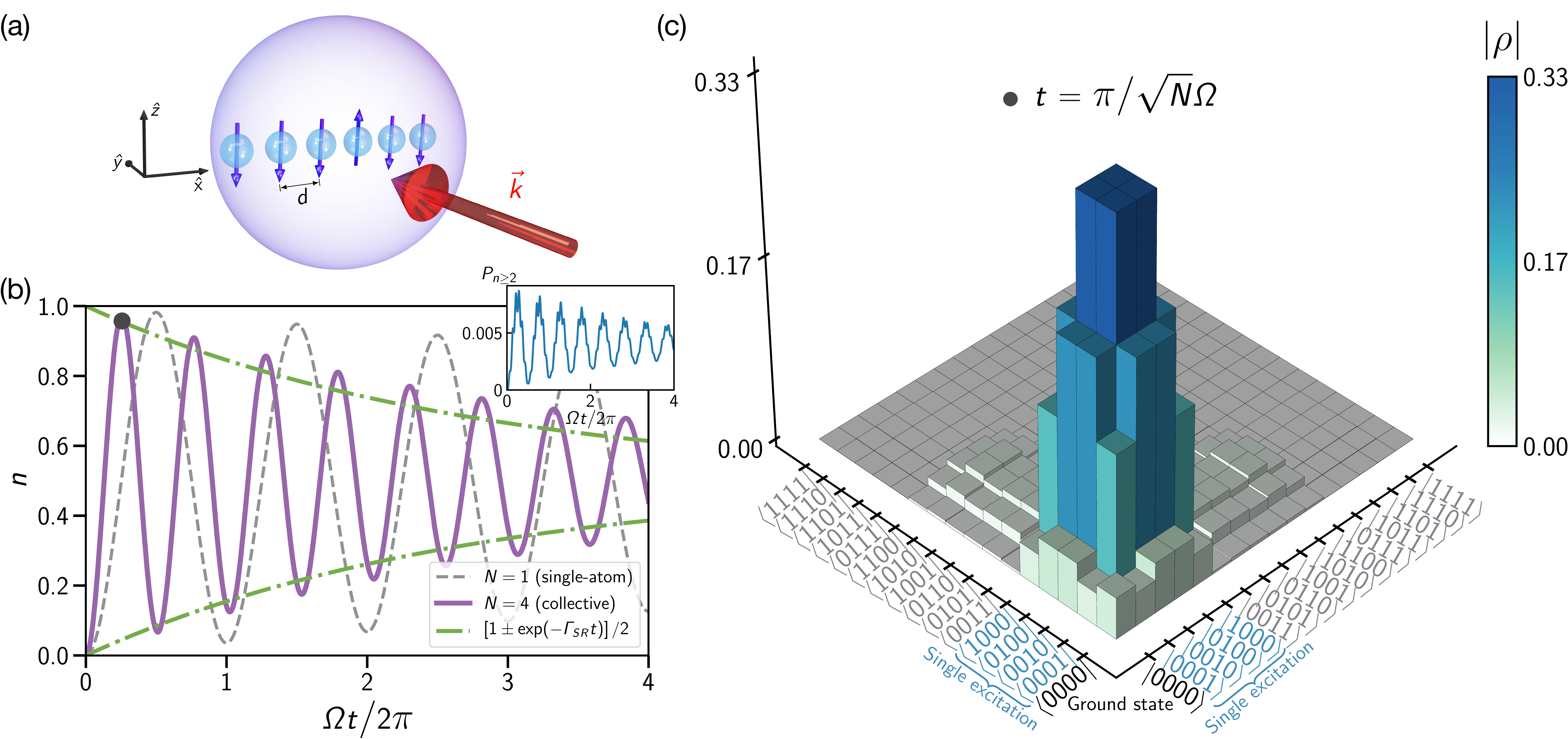}
\caption{(a) Regular chain of two-level atoms with spacing $d$ along $\hat x$ and incident laser of wave-vector $\mathbf{k}=k\hat y$, polarized along $\hat z$. The strong dipole-dipole interaction generates a large energy shift of the superradiant state, which permits the blockade effect (purple sphere). (b) Dynamics of the excited population $n$ for $N=1$ (dashed grey line) and $N=4$ atoms (purple line), for a driving Rabi frequency $\Omega/\Gamma=65$. For the $N=1$ case, the atom is driven at resonance ($\Delta=0$). For $N=4$, the detuning is chosen to drive the superradiant mode ($\Delta=\Delta_\mathrm{SR}$), and the system is composed of a regular chain with spacing $kd=0.1$. The inset shows the probability of having many excitations to be $\lesssim 10^{-2}$ throughout the evolution. (c) Density matrix for a chain with $N=4$ atoms, with spacing $kd=0.1$, driven by a laser with detuning $\Delta/\Gamma=649$, matching the superradiant-mode shift, and with a Rabi frequency $\Omega/\Gamma=65$. \label{Fig:CRO_DM}}
\end{figure*}

We consider a linear chain of $N$ two-level atoms excited with a linear polarization orthogonal to the chain, so the induced electric dipoles couple through the exchange of real and virtual photons. Their dynamics is described (under both Markov and rotating-wave approximation) by a quantum master equation of the form ($\hbar \equiv 1$)~\cite{Stephen1964,Lehmberg1970,Friedberg1973}
\begin{align}
\frac{d}{dt} \hat \rho = -\mi [\hat H, \hat \rho] + \mathcal{L}(\hat \rho),
\label{eq:meq}
\end{align}
where the coherent Hamiltonian dynamics is given by
\begin{align}
\hat H &= - \Delta \sum_i  \hat \sigma_i^+  \hat \sigma_i^- +  \frac{1}{2} \sum_i \left( \Omega e^{i\mathbf{k}\cdot \mathbf{r}_i} \hat \sigma_i^+  + h.c. \right)  \nonumber \\
&+\sum_{i,j} \Delta^{ij} \hat \sigma_i^+ \hat \sigma_j^-, \label{eq:H_dd}
\end{align} 
whereas the Lindbladian part reads
\begin{align}
\mathcal{L}(\hat \rho) &= \sum_{i,j} \Gamma^{ij} \left(\hat \sigma_i^- \hat \rho \hat \sigma_j^+ - \{ \hat \sigma_j^+ \hat \sigma_i^-,\hat \rho \}\right) \label{eq:diss}. 
\end{align}
We assume that our system is being pumped by a laser with Rabi frequency $\Omega$ and detuned by $\Delta$ from $\omega_0$. The dipole-dipole nature of the interaction is embedded in the assumption of point-like dipoles whose associated Green's tensor is given by $\mathbf{G}_{ij}\equiv\mathbf{G}(\mathbf{r}_{ij})=\frac{3\Gamma}{4}\frac{e^{ikr_{ij}}}{(kr_{ij})^3}\Big[\left(k^2r_{ij}^2+ikr_{ij}-1\right)\mathds{1}_3-\left(k^2r_{ij}^2+i3kr_{ij}-3\right)\frac{\mathbf{r}_{ij}\mathbf{r}_{ij}^{T}}{r_{ij}^2}\Big]$ for $i\neq j$, where $\mathbf{r}_{ij}\equiv \mathbf{r}_i - \mathbf{r}_j$, and $\mathbf{G}_{ii}=i\frac{\Gamma}{2}\mathds{1}_3$ for the single-atom term, with $\Gamma =d_0^2k^3/3\pi\epsilon_0\hbar$ the single-atom spontaneous decay rate, $\epsilon_0$ free space permittivity, $d_{0}$ the transition dipole moment and $k=\omega_0/c=2\pi/\lambda$ its wavenumber. The elastic and inelastic terms of the dipolar interaction can be written as $\Delta^{ij}\equiv\hat{\epsilon_i}^{*}\cdot \mathrm{Re}\left\{\mathbf{G}_{ij}\right\}\cdot \hat{\epsilon_j}$ and $\Gamma^{ij}\equiv \hat{\epsilon_i}^{*}\cdot \mathrm{Im}\left\{\mathbf{G}_{ij}\right\}\cdot\hat{\epsilon_j}$, where $\hat{\epsilon_i}$ is the polarization of the $i$-th dipole, which we choose to be $\hat\epsilon_i=\hat \epsilon=\hat z$. We have here considered a regular chain of atoms of spacing $d$ along $\hat x$, with an incident laser propagating along $\mathbf{k}=k\hat y$ and polarized along $\hat z$ (see Fig.~\ref{Fig:CRO_DM}a). In this configuration, the atoms couple only through the $z$-polarization.

Collective effects have been studied extensively in the low-excitation limit (linear-optics) regime, with reports on collective frequency shifts, sub- and superradiance \cite{Bienaime2012, Pellegrino2014,Araujo2016, Guerin2016,Roof2016,Guerin2017}. The superradiant (SR) state in this context, also labelled \textit{timed Dicke state}~\cite{Scully2006}, can be identified precisely by diagonalizing the single-excitation (linear-optics) coupling matrix $M^{ij}\equiv\Gamma^{ij}+i\Delta^{ij}$~\cite{Goetschy2011,Goetschy2011b,Skipetrov2014,Bellando2014}: it corresponds to the eigenstate whose eigenvalue has the largest real part, $\Gamma_{SR}$, while its imaginary part corresponds to its energy $\Delta_{SR}$ with respect to the atomic transition~\cite{doEspiritoSanto2020}. 

The generalization of the protocol presented in Fig.~\ref{Fig:scheme}b to $N>2$ consists in addressing a single superradiant eigenstate of the interacting Hamiltonian. In our setup, this is achieved by a homogeneous in-phase illumination by the drive and matching the laser detuning precisely to the interaction shift, $\Delta=\Delta_{SR}$. A significant blockade effect requires a large energy shift, thus a small lattice spacing, and a Rabi frequency larger than the decoherence rate ($\Omega > \Gamma_{SR}$). Here we chose $\Omega=0.1\Delta_{SR}$.

The resulting dynamics is illustrated in Fig.~\ref{Fig:CRO_DM}b for the case of $N=4$ atoms spaced by $d=0.1\,k^{-1}$, where we show the total excited population
$n=\sum_{i=1}^N \left<\hat n_i\right>$, with $\hat n_i = \hat\sigma_i^+\hat\sigma_i^-$, as a function of time. We first notice that $n$ is always less than unity and the inset plot confirms that the probability $P_{n\geq 2}$ of having more than one excitation in the system is very small ($\lesssim 10^{-2}$). The amplitude of these oscillations is damped by an exponential envelope (dot-dashed green lines) consistent with the dominating decay rate $\Gamma_{SR}$, indicating that we are addressing specifically the SR state. Secondly, we observe that the frequency of oscillation is twice as large as the one of a driven single-atom (orange dashed line), which is thus consistent with the $\sqrt{N}\Omega$ scaling of the collective Rabi oscillations. 

Furthermore, the blockade effect is confirmed by examining the time evolution of the density matrix: the dynamics is the one of an effective two-level system composed of the ground-state and the SR state. At time $t=\pi/\sqrt{N}\Omega$, after switching on the pump, we reach a state with the largest probability of finding a single excitation, with large off-diagonal coherences, as shown in Fig.~\ref{Fig:CRO_DM}c. This can be confirmed by a fidelity of $\mathcal{F}=\bra{SR}\hat\rho\ket{SR}=0.96$ with the SR state. Note that in our system the SR state is not equivalent to the symmetric, $W$-state $\ket{W}\equiv(1/\sqrt{N})\sum_{i=1}^{N}\ket{g...ge_ig...g}$ due to boundary conditions. This analysis, combined with the fact that the multi-excitation states are practically not populated (see inset of Fig.~\ref{Fig:CRO_DM}b) confirms that the blockade regime is achieved.

We remark that our protocol relies on the sufficiently large energy shift of the superradiant state, and a proper addressing of that state with the pump laser. In our case, large shifts are made possible by our choice of a subwavelength lattice spacing ($kd\sim 0.1$), allowing the near-field $1/r^3$ terms to dominate the dipolar interactions. This shift could be further enhanced using, e.g., 2D geometries, where the number of neighbouring atoms is higher. Furthermore, our geometry couples dipoles with a polarization orthogonal to the atomic chain (see Fig.~\ref{Fig:CRO_DM}), yet the coupling could be tuned turning the linear polarization around $\mathbf{k}$, as encoded in the Green's tensor $\mathbf{G}_{ij}$. For example, the near-field terms cancel at the ``magic angle''  $\theta = \sin^{-1} (1/\sqrt{3})$ between the chain axis and the polarization, and take an opposite sign below this value. Finally, it is interesting to note that boundary effects could be reduced using systems with a higher degree of symmetry, such as rings~\cite{Cremer2020}.

\begin{figure}[!h]
\centering
 \includegraphics[width=\columnwidth]{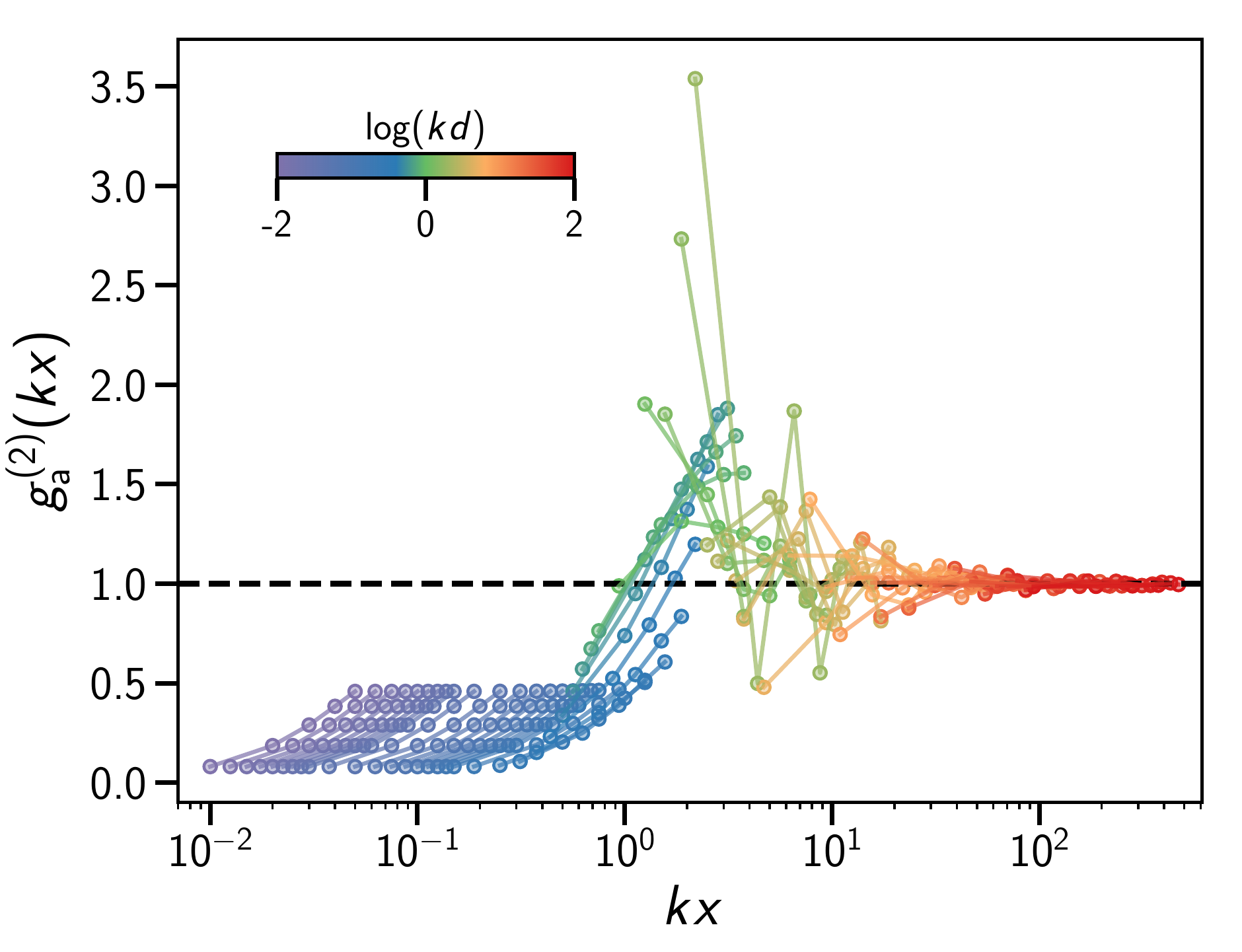}
\caption{Atomic pair correlation $g^{(2)}_a$ as a function of the lattice step $x$ for chains of $N=6$ atoms apart by different values of lattice spacing $d$. \label{Fig:g2_pair}}
\end{figure}

We now proceed to discuss a blockade radius, which describes the distance within which atoms are expected to mutually induce a blockade of excitations. To this end, we introduce an average atomic pair correlation function \cite{Ates2006, Labuhn2016}
\begin{equation}
    g^{(2)}_a(s)=\frac{1}{N_s}\sum_{i}\frac{\langle \hat{n}_i\hat{n}_{i+s}\rangle}{\langle \hat{n}_i\rangle\langle \hat{n}_{i+s}\rangle},
\end{equation}
where $N_s$ corresponds to the number of sites in the chain that are distant by $s$ sites. As can be observed in Fig.~\ref{Fig:g2_pair}, $g^{(2)}_a$ is below unity for atoms closer than $1/k$, indicating that excitations are blocked; it increases above unity as the sample size is of order of the wavelength, presenting an anti-blockade (also known as excitation-facilitating) regime \cite{Ates2007,Weber2015,Letscher2017,Kara2018}. This allows to define a blockade radius of $\sim 1/k$ for induced dipole interactions.


While probing directly the states of a subwavelength sample remains a huge challenge, photon blockade also leaves a direct signature on the scattered light. Apart from the collective Rabi oscillations discussed before, its signature can be found in the normalized intensity-intensity correlations of light~\cite{Eloy2018}
\begin{equation}\label{eq:g2_l}
g^{(2)}_l(\tau)\equiv \lim_{t\rightarrow \infty}\frac{\langle \hat{E}^{-}(t)\hat{E}^{-}(t+\tau)\hat{E}^{+}(t+\tau)\hat{E}^{+}(t)\rangle}{\langle \hat{E}^{-}(t)\hat{E}^{+}(t) \rangle^2},
\end{equation}
where we have used normal ordering for $g^{(2)}_l$. The electric field was computed using the far-field expression for the scattered field $\hat{E}^{+}\sim \sum_{j=1}^{N}e^{-ik\hat{n}.\mathbf{r}_j}\hat{\sigma}^{-}_j$, in the direction $\hat{n}=\hat{y}$ where it has the polarization $(\hat{n}\times\hat\epsilon)\times\hat{n}=\hat{z}$ of the incident laser. Here the contribution of the incident laser has been discarded: this corresponds to observing the emitted light a few degrees off the laser axis, which yields similar results since we are considering subwavelength samples.

\begin{figure}[ht!]
\centering
\includegraphics[width=\columnwidth]{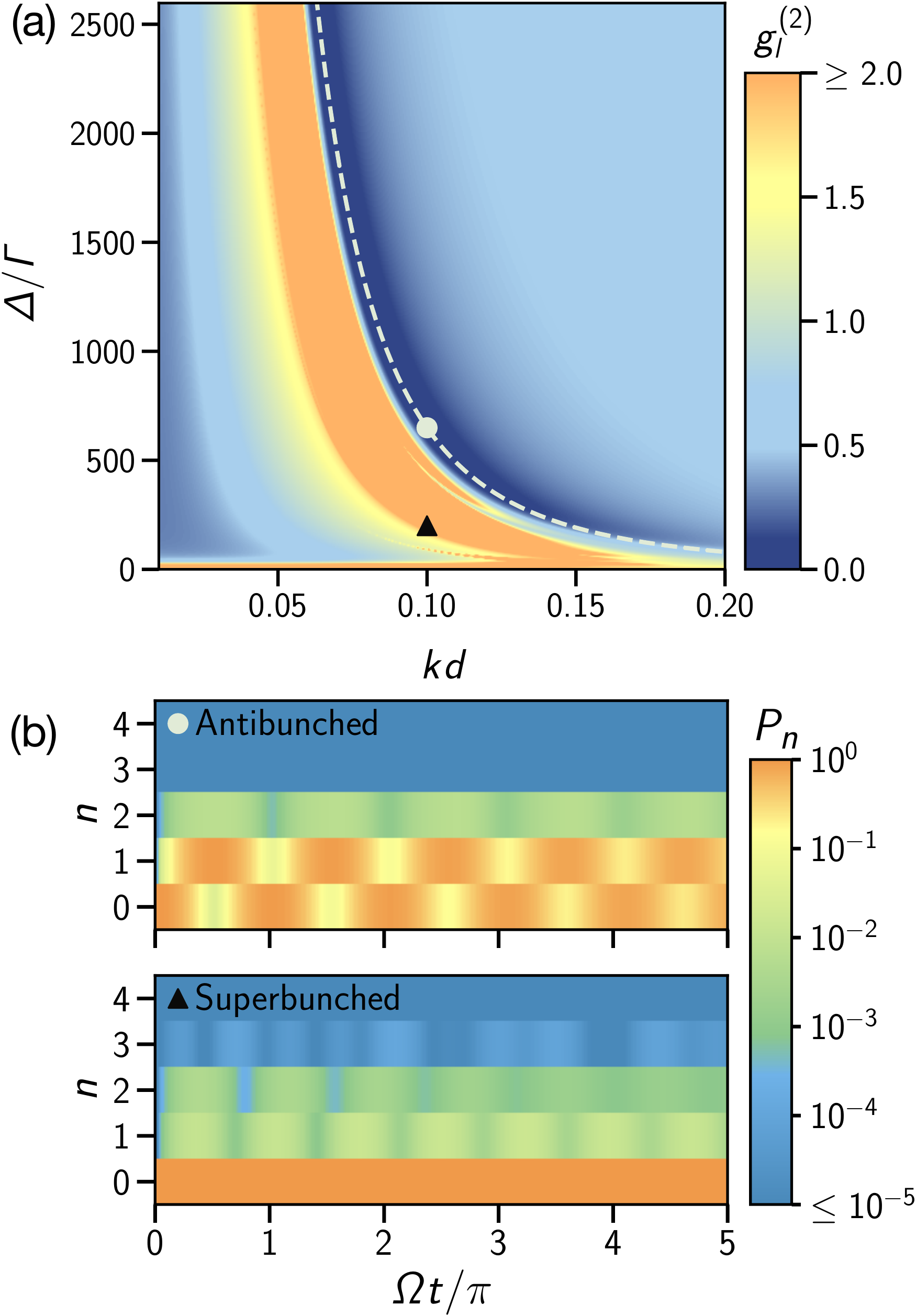}
\caption{(a) Map of second-order correlation function $g_l^{(2)}(0)$ for the light field scattered by a chain with $N=4$ as a function of detuning $\Delta$ and spacing $kd$. The strongly antibunched states lie on top of the dark-blue area, which closely coincides with the single-excitation, superradiant energy shift (dashed white line \cite{SupMat}) for each spacing $kd$. (b) Probability $P_n$ of exciting $n$ atoms as a function of time for two cases highlighted in (a) (white dot and black triangle, respectively): an antibunched (photon-blockaded) state (top) with $g_l^{(2)}(0)=0.03$ and a superbunched state (bottom) with $g^{(2)}_l(0)=5.13$. \label{Fig:g2}}
\end{figure}

The light statistics describe photon bunching when $g_l^{(2)}(0)>1$, superbunching for $g_l^{(2)}(0)>2$ and antibunching (or photon blockade) when $g_l^{(2)}(0)<1$ \cite{Walls2007} --- the latter necessarily implying nonclassicality \cite{Kimble1977}. A map of the $g_l^{(2)}(0)$ is presented in Fig.~\ref{Fig:g2}a: we have checked that the values much smaller than unity correspond to the excitation-blockade regime \cite{SupMat}, where the system oscillates between ground and single-excitation states~\cite{Kimble1977}. This map shows with which driving frequency the SR state should be addressed for different lattice spacing (see dark blue region corresponding to the antibunching of photons and  dashed white line describing the superradiant energy shift \cite{SupMat}). The detuning required to reach the photon-blockade region follows a $1/(kd)^3$ decay, as one expected from the dominating term of the dipolar interactions at short distances.

The interactions that allow for the blockade regime generate other kinds of collective states. This is already visible from the analysis of the $g_l^{(2)}(0)$ in Fig.~\ref{Fig:g2}a, where superbunching is observed (orange region). To illustrate this point, the probability $P_n$ of exciting $n$ atoms is monitored as a function of time: the blockade regime is characterized by a population which oscillates between the ground and single-excitation states (see Fig.~\ref{Fig:g2}b). The probability to explore a multi-excited state is comparatively small, and in phase with the single-excitation state, which suggests there are off-resonant events directly resulting from the SR state excitation. In contrast, driving the system at a different frequency can lead to a weakly-excited ($P_0\approx 1$) yet superbunched state (see Fig.~\ref{Fig:g2}a), where the single- and double-excitation states have comparable probabilities to be explored~\cite{SupMat}. 

In conclusion, we have shown photon blockade using ground-state neutral atoms in subwavelength lattices by addressing a collective SR state of the system, whose energy can be shifted far away from the atomic resonance by induced dipole-dipole interactions. This protocol can be in principle implemented using atoms with long-lived clock transitions, which are considerably less sensitive to stray magnetic and electric fields than Rydberg atoms, offering robustness for coherent control. Note that while we have here used the SR state, a wealth of collective states is actually generated by the dipole-dipole interactions, whose properties remain to be analyzed and possibly harnessed. In this context, the recent progresses in the experimental realization of subwavelength optical lattices holds many promises for creating and manipulating highly-entangled states \cite{Subhankar2019,Tsui2019,Rui2020} and opens the way for a possible implementation of the proposal discussed in this work. Note that our proposal is not limited to atomic setups, but also for other platforms being able to implement large dipole-densities at subwavelength scales, such as with color centers~\cite{Juan2016,Bradac2017,Rain2018}.

\begin{acknowledgments}
We thank Ana Maria Rey and Tommaso Macr\`i for helpful discussions.  A.C. and R.B. are supported by FAPESP through Grants No. 2017/09390-7, 2018/01447-2 and 2018/15554-5. R.B. received support from the National Council for Scientific and Technological Development (CNPq) Grant Nos. 302981/2017-9 and 409946/2018-4. J.S.~is supported by the French National Research Agency (ANR) through the Programme d'Investissement d'Avenir under contract ANR-11-LABX-0058\_NIE within the Investissement d'Avenir program ANR-10-IDEX-0002-02. Part of this work was performed in the framework of the European Training Network ColOpt, which is funded by the European Union (EU) Horizon 2020 programme under the Marie Sklodowska-Curie action, grant agreement No. 721465. R. B. and R. K. received support from project CAPES-COFECUB (Ph879-17/CAPES 88887.130197/2017-01).
\end{acknowledgments}

\bibliography{BiblioCollectiveScattering.bib}

\appendix

\newpage
\setcounter{equation}{0}
\setcounter{figure}{0}
\setcounter{table}{0}
\makeatletter
\renewcommand{\theequation}{S\arabic{equation}}
\renewcommand{\thefigure}{S\arabic{figure}}
\renewcommand{\bibnumfmt}[1]{[S#1]}
\renewcommand{\citenumfont}[1]{S#1}

\onecolumngrid
\vspace{0.5cm}
\begin{center}
	\textbf{\large Supplementary Material: Photon blockade with ground-state neutral atoms}\\[.2cm]
\end{center}	

\section{Energy shift of the single-excitation symmetric state for $N=4$}

We here discuss the analytical calculation of the energy shift $\Delta_{SR}$ of the superradiant state for $N=4$, used to compute its dependence on the lattice spacing  $kd$ in the $g^{(2)}_l$ density map (see Fig.~\ref{Fig:g2} in the main text). 
Writing the virtual photon exchange term $\Delta^{ij}$ (see Eq.~\ref{eq:H_dd} in the main text) in the single-excitation manifold (using the canonical basis $\left\{\ket{0,0,0,1},\ket{0,0,1,0},\ket{0,1,0,0},\ket{1,0,0,0}\right\}$) leads to the following coupling matrix
\begin{equation}
	\Delta^{ij}_{\mathrm{1e}} = 
	\begin{pmatrix}
		0 & \Delta_1 & \Delta_2 & \Delta_3 \\
		\Delta_1 & 0 & \Delta_1 & \Delta_2 \\
		\Delta_2  & \Delta_1  & 0 & \Delta_1  \\
		\Delta_3 & \Delta_2 & \Delta_1 & 0
	\end{pmatrix},
\end{equation}
where $\Delta_s$ refers to the interaction between atoms apart by $s$ sites
\begin{equation}
	\Delta_s=-\frac{3\Gamma}{4}\frac{1}{(skd)^3}\big[\left((skd)^2-1\right)\cos\left(skd\right)-skd\sin\left(skd\right)\big].
\end{equation} 
Diagonalizing this matrix allows us to identify the most superradiant state, whose energy shift reads
\begin{equation}\label{eq:Delta_SR_analytical}
	\begin{aligned}
		\Delta_{SR} = \frac{1}{2}\Big(\Delta_1+\Delta_3
		+\sqrt{5\Delta_1^2+8\Delta_1\Delta_2+4\Delta_2^2-2\Delta_1\Delta_3+\Delta_3^2} \Big).
	\end{aligned}
\end{equation}

\section{Number of excitations}

\begin{figure*}[!h]
	\centering
	\includegraphics[width=\columnwidth]{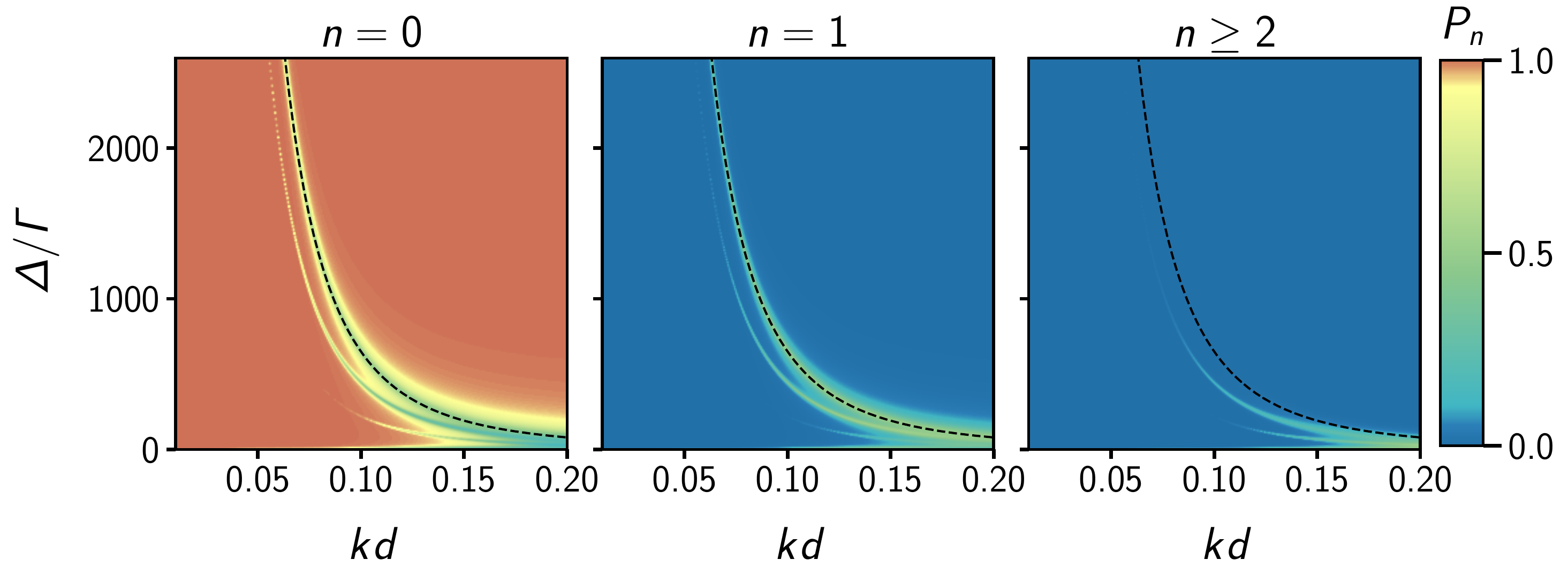}
	\caption{ Probability $P_n$ of finding $n$ excited atoms as a function of detuning $\Delta$ and spacing $d$, for $N=4$. The dashed black line corresponds to the analytical $\Delta_{SR}$, as calculated in Eq.~\ref{eq:Delta_SR_analytical}.\label{Fig:pn_map}}
\end{figure*}

In Fig.~\ref{Fig:pn_map} we plot the probability $P_n$ of finding $n$ excited atoms in the steady-state, for a chain of $N=4$ atoms and as a function of the detuning $\Delta$ and the lattice spacing $d$. In the region where the photon antibunching (i.e., $g_l^{(2)}(0)<1$) is observed (see Fig.~\ref{Fig:g2}a and the associated discussion in the main text), the probability $P_{n\geq 2}$ to get two or more excitations is very close to zero, while the system shows equal probability of being in the ground state or in a single-excited state. The other regions with a significant single-excitation population $P_{n=1}$ are also characterized by a significant many-excitation population $P_{n\geq2}$, which suggests the presence of other strongly-shifted collective states, yet which do not present a photon-blockade effect.

\end{document}